\documentclass[12pt,reqno]{article}
\usepackage{amssymb}
\usepackage{amsthm}
\usepackage{amsmath}
\usepackage{multirow}
\usepackage{float}
\usepackage{graphicx}
\usepackage[font=small,labelfont=bf]{caption}
\usepackage{listings}
\newtheorem{theorem}{Theorem}[section]

\newtheorem{lemma}[theorem]{Lemma}
\newtheorem{proposition}[theorem]{Proposition}

\newtheorem{definition}[theorem]{Definition}
\newtheorem{example}[theorem]{Example}

\newtheorem{remark}[theorem]{Remark}

\begin{document}
\begin{center} \Large{\bf  A Novel Method to Generate Key-Dependent S-Boxes with Identical Algebraic Properties}
\end{center}
\medskip

\begin{center}
Ahmad Y. Al-Dweik$^*$, Iqtadar Hussain$^*$, Moutaz Saleh$^{**}$, M. T. Mustafa$^*$\\

{$^{*}$Department of Mathematics, Statistics and Physics, Qatar
University, Doha, 2713, State of Qatar}

{$^{**}$Department of Computer Science and Engineering, Qatar University, Doha,
2713, State of Qatar}

aydweik@qu.edu.qa, iqtadarqau@qu.edu.qa, moutaz.saleh@qu.edu.qa
and tahir.mustafa@qu.edu.qa
\end{center}
\begin{abstract}
The s-box plays the vital role of creating confusion between the ciphertext and secret key in any cryptosystem, and is the only nonlinear component in many block ciphers. Dynamic s-boxes, as compared to static, improve entropy of the system, hence leading to better resistance against linear and differential attacks. It was shown in \cite{27} that while incorporating dynamic s-boxes in cryptosystems is sufficiently secure, they do not keep non-linearity invariant. This work provides an algorithmic scheme to generate key-dependent dynamic $n\times n$ clone s-boxes having the same algebraic properties namely bijection, nonlinearity, the strict avalanche criterion (SAC), the output bits independence criterion (BIC) as of the initial seed s-box. The method is based on group action of symmetric group $S_n$ and a subgroup $S_{2^n}$ respectively on columns and rows of Boolean functions ($GF(2^n)\to GF(2)$)  of s-box.  Invariance of the bijection, nonlinearity, SAC, and BIC for the generated clone copies is proved. As illustration, examples are provided for $n=8$ and $n=4$ along with comparison of the algebraic properties of the clone and initial seed s-box. The proposed method is an extension of \cite{n15,n16,n30,n32} which involved group action of  $S_8$ only on columns of Boolean functions ($GF(2^8)\to GF(2)$ ) of s-box. For $n=4$, we have used an initial $4\times 4$ s-box constructed by Carlisle Adams and Stafford Tavares \cite{Adms1990} to generated $(4!)^2$ clone copies. For $n=8$, it can be seen \cite{n15,n16,n30,n32} that the number of clone copies that can be constructed by permuting the columns is $8!$. For each column permutation, the proposed method enables to generate $8!$ clone copies by permuting the rows. 
\end{abstract}
\section{Introduction}
Cryptography has emerged as a key solution for protecting
information and securing data transmission against passive and
active attacks. Substitution box or s-box is a vital component of
symmetric block encryption schemes such as Data Encryption
Standard (DES), Advanced Encryption Standard (AES) and
International Data Encryption Algorithm (IDEA). The cryptographic
strength of these encryption systems mainly depends upon the
efficiency of their substitution boxes being the only components
capable of inducing the nonlinearity in the cryptosystem \cite{1}.
This attracted the attentions of many researches to design
cryptographically potent s-boxes for the sake of developing robust
encryption schemes. Theoretically, there are several properties
that can evaluate the performance of a proposed s-box \cite{2}.
The most commonly applied properties are the  bijective property,
nonlinearity, strict avalanche criteria (SAC) and bits
independence criterion (BIC).

Depending on the design nature of s-box, it can be classified into
either static or dynamic. The static s-box is one whose values are
key-independent and once defined by the designer it is maintained
during the whole encryption process. This means that the same
s-box will be used in every round, and so it might be vulnerable
to cryptanalysis. On the other hand,  dynamic s-boxes do not
suffer from fixed structure block ciphers since the s-boxes itself
are changed in every encryption round and it is considered
key-dependent. Hence, the adoption of dynamic s-boxes improves the
security of the system and better resists against various
differential and cryptanalysis attacks \cite{3}.

Many researchers have explored several ideas for s-box design such
as randomness, dynamicity, and key-dependency. For instance,
Krishnamurthy and Ramaswamy employed s-box rotation and used it as
an additional component in the traditional AES algorithm to design
a dynamic s-box \cite{4}. The process consists of three steps in
which the s-boxes are rotated based on fixed, partial and whole
key values to increase their security. In \cite{5}, Piotr
Mroczkowski proposed an algorithm to replace the available s-boxes
through using pseudo-randomly generators to design similar s-boxes
in both encryption and decryption processes. The work claims that
changing of s-boxes could prevent intruders from receiving enough
information to execute effective cryptanalysis attack. Stoianov
\cite{6} proposed a novel approach for changing the s-boxes used
in the AES algorithm through introducing two new s-boxes known as
SBOXLeft and S-BOXRight that employs the left and right diagonals
as the axis of symmetry.

The practice of using key-dependent generated s-boxes in
cryptography has been also extensively studied in the literature.
For instance, the work in \cite{7} used RC4 algorithm to generate
key dependent s-boxes based on the input key. The authors showed
that their generated dynamic s-boxes have increased the AES
complexity and also make the differential and linear cryptanalysis
more difficult. In \cite{8}, Kazlauskas et al. proposed an
approach to randomly generate key dependent s-box that rely on
changing only one bit of the secret key. Their approach is claimed
to solve the problem of the fixed structure s-boxes, and increase
the security level of the AES block cipher system due to its
resistance of linear and differential cryptanalysis attacks. Ghada
Zaibi et al. presented dynamic s-boxes based on one dimensional
chaotic maps and evaluated its efficiency compared to the static
s-box \cite{9}. Their findings showed that AES using dynamic
chaotic s-box is more secure and efficient than AES with static
s-box. In their work \cite{10}, Jie Cui et al. proposed to
increase the complexity and security of AES s-box by modifying the
affine transformation cycle. The evaluation results suggested that
the improved AES s-box has better performance and can readily be
applied to AES. Likewise, Anna Grocholewska-Czurylo \cite{11}
described an AES-like dynamic s-boxes generated using finite field
inversion. The significant remark for this work indicates that
removing the affine equivalence cycles from s-boxes does not
influence on their cryptographic properties. Julia Juremi et al.
\cite{12} proposed a key-dependent s-box to enhance the security
of AES algorithm through employing a key expansion algorithm
together with s-box rotation. The obtained results showed that the
enhancement on the original AES does not violate the security of
the cipher. Similar to the work in \cite{8}, Razi Hosseinkhani et
al. \cite{13} introduced an algorithm to generate dynamic s-box
from cipher key. The quality of this algorithm was tested by
changing only two bits of cipher key to generate new s-boxes. The
authors claim that the key advantage of this algorithm is that
various s-boxes can be generated by changing cipher key.
 Iqtadar Hussain et al. \cite{n15} presented a method for constructing $8\times8$ s-boxes using the Liu J substitution box as a seed during the creation process. The proposed design relies on the symmetric group permutation operation which is embedded in the algebraic structure of the new s-box.  An extension of the above work was conducted by the same authors in \cite{n16}. They proposed a novel method that uses the symmetric group permutation based on the characteristics of affine-power-affine structure to generate nonlinear s-box component with the possibility to incorporate 40320 unique instances. The work presented a deep analysis to evaluate the properties of these new s-boxes and determine its suitability to various encryption applications.

In the middle of the last decade, several attempts were made to
design robust dynamic s-boxes for symmetric cryptography systems.
For instance, Oleksandr Kazymyrov et al. \cite{14} described an
improved method based on the analysis of vectorial Boolean
functions properties for selection of s-boxes with optimal
cryptographic properties that would lead to provide high level of
robustness against various types of attacks. Mona Dara et al.
\cite{15} used chaotic logistic maps with cipher key to construct
key dependent s-boxes for AES algorithm. The proposed s-box was
tested against equiprobable input/output XOR distribution, key
sensitivity, nonlinearity, SAC and BIC properties. In \cite{16},
Eman Mahmoud et al. designed and implemented a dynamic AES-128
with key dependent s-boxes using pseudo random sequence generator
with linear feedback shift Register. The quality of the
implemented s-boxes is experimentally investigated, and compared
with original AES in terms of security analysis and simulation
time. In their work \cite{17}, Sliman Arrag et al. improved s-box
complexity through using nonlinear transformation algorithm.
Further, they also adjusted key expansion schedule and use s-box
lookup table to make it dynamic. Fatma Ahmed et al. \cite{18}
proposed s-boxes by using dynamic key and employed it as a
repository for randomly selecting s-boxes in AES algorithm.

Using pseudo-random generators have also been broadly employed to
design dynamic key-dependent s-boxes.   Following the approaches
in \cite{5,8,16}, Adi Reddy et al. \cite{19} enhanced the AES
security by designing s-boxes using random number generator for
sub keys in key expansion module of their algorithm. The work
showed that the proposed s-boxes are free from linear and
differential cryptanalysis attack, and also it required less
memory with high processing speed compared to other existing
improvements. In \cite{20}, Kazlauskas et al. modified the
existing AES algorithm by generating key-dependent s-boxes using
random sequences. The authors claim that the new generated
algorithm outperform the traditional AES. Balajee Maram et al.
\cite{21} generated key-dependent s-boxes by using Pseudo-Random
generator. Their statistical analysis shows that the proposed
algorithm could generate s-boxes faster than other available
algorithms.

Recently, Shishir Katiyar et al. \cite{22} generated dynamic
s-boxes by using logistic maps. The efficiency of the proposed
dynamic s-box was reviewed and analyzed over static s-box. The
carried out experiments have shown that the key-dependent s-box
satisfies all the cryptographic properties of good s-box and can
enhance the security due to its dynamic nature. In \cite{23},
Tianyong Ao et al. made affine transformation key-dependent to
generate dynamic s-boxes for their algorithm. The authors
investigations revealed that the algebraic degree of an s-box is
conditional invariant under affine transformation. Unal C. et al.
\cite{24} proposed a secure image encryption algorithm design
using dynamic chaos-based s-box. The work showed that the
developed s-box based image encryption algorithm is secure and
speedy. In \cite{25}, Agarwal P. et al. developed a key-dependent
dynamic s-boxes using dynamic irreducible polynomial and affine
constant. This latter algorithm was used by Amandeep Singh et al.
to \cite{26} develop a new dynamic AES in which s-boxes are made
completely key-dependent.
In \cite{n30}, Iqtadar Hussain et al. proposed an encryption algorithm based on the substitution-permutation performed by the S8 Substitution boxes and also incorporates three different chaotic maps. The presented simulation and statistical results showed that the proposed encryption scheme is secure against different attacks and resistant to the channel noise.

Despite the extensive works by many researcher towards designing
key-dependent s-boxes, Chuck Easttom \cite{27} showed that while
key-dependent variations of Rijndael are sufficiently secure, they
do not demonstrate improved non-linearity over the standard
Rijndael s-box, instead they do introduce additional processing
overhead. 
To address this claim, Amir Anees et al. \cite{n32} proposed a new method for creating multiple substitution boxes with the same algebraic properties using permutation of symmetric group on a set of size 8 and bitwise XOR operation. Their analysis demonstrated that the proposed substitution boxes can resist differential and linear cryptanalysis and sustain algebraic attacks. 
Ultimately to further extend the latter work, we propose a novel method to generate key dependent s-boxes with identical algebraic properties by applying two permutations on both of  the inputs and
outputs vectors of an initial s-box. A rigorous analysis is also presented to evaluate the properties of the newly created s-boxes particularly the bijection, nonlinearity, SAC, and BIC invariant.

The remainder of this paper is organized into following sections:
Section 2 discusses in details the common algebraic properties of
the s-box, Section 3 presents main theorem and describes the
proposed key-dependent dynamic s-box generation algorithm. The conclusion is provided in Section 4.
\section{Preliminaries}
A Boolean function of $n$ inputs, $f(x_1, x_2,...,x_n)$, is a
function of the form $f:\{0,1\}^n\rightarrow \{0,1\}$. It can be
regarded as a binary vector $\textbf{f}$ of length $2^n$,
 where $\textbf{f}$ is the rightmost column of the truth table describing this
 function. We denote the set of all Boolean functions of $n$ inputs $B_n$.

The Boolean functions can serve as the $n$ output bits of the
s-box. Let $f_1, f_2 ..... f_n$ be the $n$ Boolean functions,
where each function $f_i$ corresponds to a binary vector
$\textbf{f}_i$ of length $2^n$. Then the s-box $S = [\textbf{f}_1,
\textbf{f}_2 ..... \textbf{f}_n]$ is a $2^n \times n$ bit matrix
with the $\textbf{f}_i$ as column vectors. Any given input vector
$x = x_1, x_2, ..., x_n$, maps to an output vector $y = y_1, y_2
,..., y_n$, by the assignment $y_i = f_i(x_1, x_2 ,..., x_n)$.

The main purpose of this paper is providing a key-dependent
algorithm that generates a set of Boolean functions $f_1, f_2
,..., f_n$, such that the corresponding s-box is bijective,
nonlinear, and fulfills SAC and BIC. Before introducing this
algorithm, let us first revise these four algebraic properties.
\subsection{Bijection}
It ensures that all possible $2^n$ $n$-bit input vectors will map
to distinct output vectors (i.e., the s-box is a permutation of
the integers from $0$ to $2^n-1$).

\begin{proposition}\label{proposition 2.1} \cite{Adms1990}
The necessary and sufficient condition for the s-box S to be
bijective is that any linear combination of the columns of S has
Hamming weight $2^{n-1}$. (i.e., $wt(a_1\textbf{f}_1\oplus
a_2\textbf{f}_2\oplus...\oplus a_n\textbf{f}_n)=2^{n-1}$, where
the $a_i\in \{0,1\}$ and the $a_i$ are not all simultaneously
zero).
\end{proposition}
\subsection{Nonlinearity}
It ensures that the s-box is not a linear mapping from input
vectors to output vectors (since this would render the entire
cryptosystem easily breakable).

The nonlinearity $N_f$ of a function $f$ is defined
\cite{Shannon1949} as the minimum Hamming distance between that
function and every linear function. (i.e., $N_f= min_{l\in L_n}
d_H(f,l)$, where $L_n$ is a set of the whole linear and affine
functions and $d_H(f,l)$  denotes the Hamming distance between $f$
and $l$)

\begin{remark}
Pieprzyk and Finkelstein \cite{Pieprzyk1988} claim that the
highest nonlinearity achievable with 0-1 balanced functions can be
calculated by the following equation
\begin{equation}
N_f  = \left\{
\begin{array}{cc}
   \sum\limits_{\frac{1}{2}(n-3)  \le i \le n-3 }2^{i+1} & for~~n=3,5,7,...~,\\
   \sum\limits_{\frac{1}{2}(n-4)  \le i \le n-4 }2^{i+2} & for~~n=4,6,8,...~.\\
\end{array}\right.
\end{equation}
\end{remark}
\begin{remark}
Carlisle Adams and Stafford Tavares \cite{Adms1990} stated that if
the $n$ Boolean functions of an s-box $S$ are nonlinear, then $S$
is guaranteed to be nonlinear at the bit level and at the integer
level.
\end{remark}
\begin{lemma}\label{lemma 2.4}\cite{Pieprzyk2013}
Let $f$ be a Boolean function over $\{0,1\}^n$, $B$ be an $n\times
n$ nonsingular matrix, and $\beta$ a constant vector from
$\{0,1\}^n$. Then the function $g(x) = f(xB\oplus \beta)$ has the
same nonlinearity as the function $f$ so $N_g = N_f$.
\end{lemma}
\subsection{Strict avalanche criterion}
SAC was introduced by Webster and Tavares \cite{Webster1986}.
Informally, an s-box satisfies SAC if a single bit change on the
input results in changes on a half of output bits. More formally,
a function $f : \{0,1\}^n \rightarrow GF(2)$ satisfies the SAC if
$f(x)\oplus f(x\oplus \gamma)$ is balanced for all $\gamma$ whose
weight is 1, (i.e., $wt(\gamma) = 1$). In other words, the SAC
characterizes the output when there is a single bit change on the
input.
\begin{theorem}\label{theorem 2.5}\cite{Pieprzyk2013}
Let $f : \{0,1\}^n \rightarrow GF(2)$ be a Boolean function and
$A$ be an $n\times n$ nonsingular matrix with entries from
$GF(2)$.  If $f(x)\oplus f(x\oplus \gamma)$ is balanced for each
row $\gamma$ of $A$, then the function $\psi(x) = f(xA)$ satisfies
the SAC.
\end{theorem}
\subsection{Bit independent criterion}
Given two Boolean functions $f_j$, $f_k$  in an s-box, if $f_j
\oplus f_k$  is highly nonlinear and meets the SAC, then the
correlation coefficient of each output bit pair may be close to 0
when one input bit is flipped. Thus, we can check the BIC of the
s-box by verifying whether  $f_j \oplus f_k$ $(j \neq k)$   of any
two output bits of the s-box meets the nonlinearity and SAC
\cite{Webster1986}.
\section{The main theorem and proposed key-dependent dynamic s-box algorithm}
\begin{definition} A permutation matrix is a matrix obtained by
permuting the rows of an $n\times n$ identity matrix according to
some permutation of the numbers $0$ to $n-1$. Every row and column
therefore contains precisely a single $1$ with $0$s everywhere
else, and every permutation corresponds to a unique permutation
matrix.
\end{definition}
\begin{lemma}\label{lemma 3.2}
Let $X$ be the identity s-box with the $2^n \times n$ bit matrix
$[\textbf{x}_1, \textbf{x}_2,...,\textbf{x}_n]$
 and  $P_1$ be a permutation matrix of size $n\times n$. There
exist a permutation matrix $Q_1$ of size $2^n\times 2^n$ such that
$Q_1 X=X P_1$.
\end{lemma}
\proof The post-multiplying of the permutation matrix $P_1$ with
the matrix $X$ to form the matrix $W_1=X P_1$  results in
permuting columns of the matrix $X$. Using proposition
\ref{proposition 2.1}, the $2^n \times n$ bit matrix $W_1$ is
bijective s-box. Therefore, converting the rows of the matrix
$W_1$ to the decimal representation provides a permutation $P_3$
of size $2^n$.

Let $Q_1$ be the permutation matrix  of size $2^n\times 2^n$
corresponding to the permutation $P_3$. Since the pre-multiplying
of the permutation matrix $Q_1$ with the matrix $X$ to form the
matrix $Q_1 X$  results in permuting rows of the matrix $X$, then
$Q_1 X=X P_1$.
\endproof
\begin{theorem}
Let $X$ be the identity s-box with the $2^n \times n$ bit matrix
$[\textbf{x}_1, \textbf{x}_2,...,\textbf{x}_n]$, $Y$ be an s-box
with the $2^n \times n$ bit matrix $[\textbf{y}_1,
\textbf{y}_2,...,\textbf{y}_n]$, $P_1, P_2$ be permutation
matrices of size $n\times n$, and $Q_1$ be the permutation matrix
of size $2^n\times 2^n$ satisfying $Q_1 X=X P_1$. The  $2^n \times
n$ bit matrix $Q_1 Y P_2$ is  new s-box with the same algebraic
properties: bijection, nonlinearity, SAC, BIC as the initial s-box
$Y$.
\end{theorem}
\proof Let the $n$ Boolean functions $f_1, f_2 ..... f_n$
correspond to the column vectors $\textbf{y}_i=\textbf{f}_i$ of
the matrix $Y$.

Since, the post-multiplying of the permutation matrix $P_2$ with
the matrix $Y$ to form the matrix $Y P_2$  results in permuting
columns of the matrix $Y=[\textbf{f}_1, \textbf{f}_2 ,...,
\textbf{f}_n]$, then it is clear that the  $2^n \times n$ bit
matrix $Y P_2$ is new s-box with the same algebraic properties:
bijection, nonlinearity, SAC, BIC as the initial s-box $Y$.
Therefore, it is enough to prove that the $2^n \times n$ bit
matrix $Q_1 Y=[Q_1\textbf{f}_1,Q_1 \textbf{f}_2 ,...,
Q_1\textbf{f}_n]$ is new s-box with the same algebraic properties:
bijection, nonlinearity, SAC, BIC as the initial s-box $Y$.

a) Bijection:

 The pre-multiplying of the permutation matrix $Q_1$
with the matrix $Y$ to form the matrix $Q_1 Y$  results in
permuting rows of the matrix $Y$. Using proposition
\ref{proposition 2.1}, $Q_1 Y$ is bijective s-box.

b) Nonlinearity and SAC:

Introduce $g_1(x), g_2(x), ..., g_n(x)$ to be the Boolean
functions defined by  $g_i(x) = f_i(x P_1)$ for $i=1,...,n$. Using
lemma \ref{lemma 2.4}, $N_{f_i} = N_{ g_i}$ for $i=1,...,n$. Also,
if the functions $f_i(x)$ satisfy the SAC for $i=1,...,n$, then
$f_i(x)\oplus f_i(x\oplus \gamma)$ is balanced for each row
$\gamma$ of $P_1$ for $i=1,...,n$. Therefore, using theorem
\ref{theorem 2.5}, the functions $g_i(x)$ satisfy the SAC for
$i=1,...,n$. Hence, the s-box $[\textbf{g}_1, \textbf{g}_2 ,...,
\textbf{g}_n]$ satisfies the SAC and has the same nonlinearity as
the initial s-box $Y=[\textbf{f}_1, \textbf{f}_2 ,...,
\textbf{f}_n]$.

Finally,  since $Y=[\textbf{f}_1(X), \textbf{f}_2(X) ,...,
\textbf{f}_n(X)]$ and $Q_1 X=X P_1$, then the s-box
\begin{equation}
\begin{array}{ll}
[\textbf{g}_1(X),\textbf{g}_2(X) ,..., \textbf{g}_n(X)]&=[\textbf{f}_1(XP_1),\textbf{f}_2(XP_1) ,...,\textbf{f}_n(XP_1)]\\
&=[\textbf{f}_1(Q_1X),\textbf{f}_2(Q_1X),...,\textbf{f}_n(Q_1X)]\\
&=[Q_1\textbf{f}_1(X),Q_1\textbf{f}_2(X),...,Q_1\textbf{f}_n(X)]=Q_1Y.\\
\end{array}
\end{equation}
c) BIC:

Assume that  the function $h_{ij}(x)=f_j(x) \oplus f_k(x)$  of any
two different output bits $f_j$ and  $f_k$ of the s-box $Y$ meets
the nonlinearity and SAC. Introduce $k_{ij}(x)$ to be the Boolean
functions defined by the function $k_{ij}(x)=h_{ij}(xP_1)$. Using
lemma \ref{lemma 2.4}, $N_{h_{ij}} = N_{k_{ij}}$ for $i\ne j$.
Similarly, using theorem \ref{theorem 2.5}, the functions
$k_{ij}(x)$ satisfy the SAC for $i\ne j$. Therefore, the function
$k_{ij}(x)=h_{ij}(xP_1)=f_j(xP_1) \oplus f_k(xP_1)=g_j(x) \oplus
g_k(x)$ meets the nonlinearity and SAC. Hence, the s-box
$[\textbf{g}_1, \textbf{g}_2 ,..., \textbf{g}_n]$ satisfies BIC.
\endproof
The proposed method for key-dependent dynamic s-boxes  consists of
permutations of the inputs and outputs vectors of an initial
s-box. The following algorithm provides s-boxes with identical
algebraic properties. The steps are summarized as follows.

1. Express the initial $n\times n$ s-box  as a vector $IS$
consisting of different $2^n$ values ranging between $0$ and
$2^n-1$.

2. Compute the matrix  $Y$ of size $2^n \times n$ by evaluating
the binary representation of the initial s-box. Similarly, compute the matrix  $X$ of size $2^n \times n$ by evaluating
the binary representation of  the identity s-box.

3. Use the key to construct two permutations $\sigma_1$ and
$\sigma_2$ of different $n$ values ranging between $0$ and $n-1$.

4. Construct the corresponding permutation matrices $P_1$ and
$P_2$ of size $n\times n$ for the two permutations $\sigma_1$ and
$\sigma_2$.

5. Compute the  matrix $W_1=X P_1$ of size $2^n\times n$.

6. Construct the corresponding permutation $P_3$  of size $2^n$ by getting back
the decimal representation of the matrix $W_1$ as a vector.

7. Construct the  permutation matrix $Q_1$ of size $2^n\times 2^n$
corresponding to the permutation $P_3$.

8. Compute the  matrix $W_2=Q_1 Y P_2$ of size $2^n\times n$.

9. Construct the dynamic key-dependent s-box by getting back the
decimal representation of the matrix $W_2$ as the vector $NS$.

10. Detect the possible fixed point and reverse fixed point of the constructed vector $NS$.  

11. In case there are fixed points or reverse fixed points, update the initial permutations using $\sigma_1=\bar\sigma_1 \sigma_1, \sigma_2=\bar\sigma_2 \sigma_2$  such that $\bar\sigma_1, \bar\sigma_2 \in S_{n}$ where $S_{n}$ is the symmetric group and then return back to step 4. Otherwise, end the algorithm and produce the vector $NS$  as clone dynamic key-dependent s-box.

An efficient Maple code implementation of the method described in this section is presented in the Appendix.
\begin{remark}
The two permutations $\sigma_1$ and
$\sigma_2$  of size $n$ can be extracted from the key using factorial number system and  Lehmer code.
\end{remark}
\begin{figure}[H]
\centering
\includegraphics[scale=0.6]{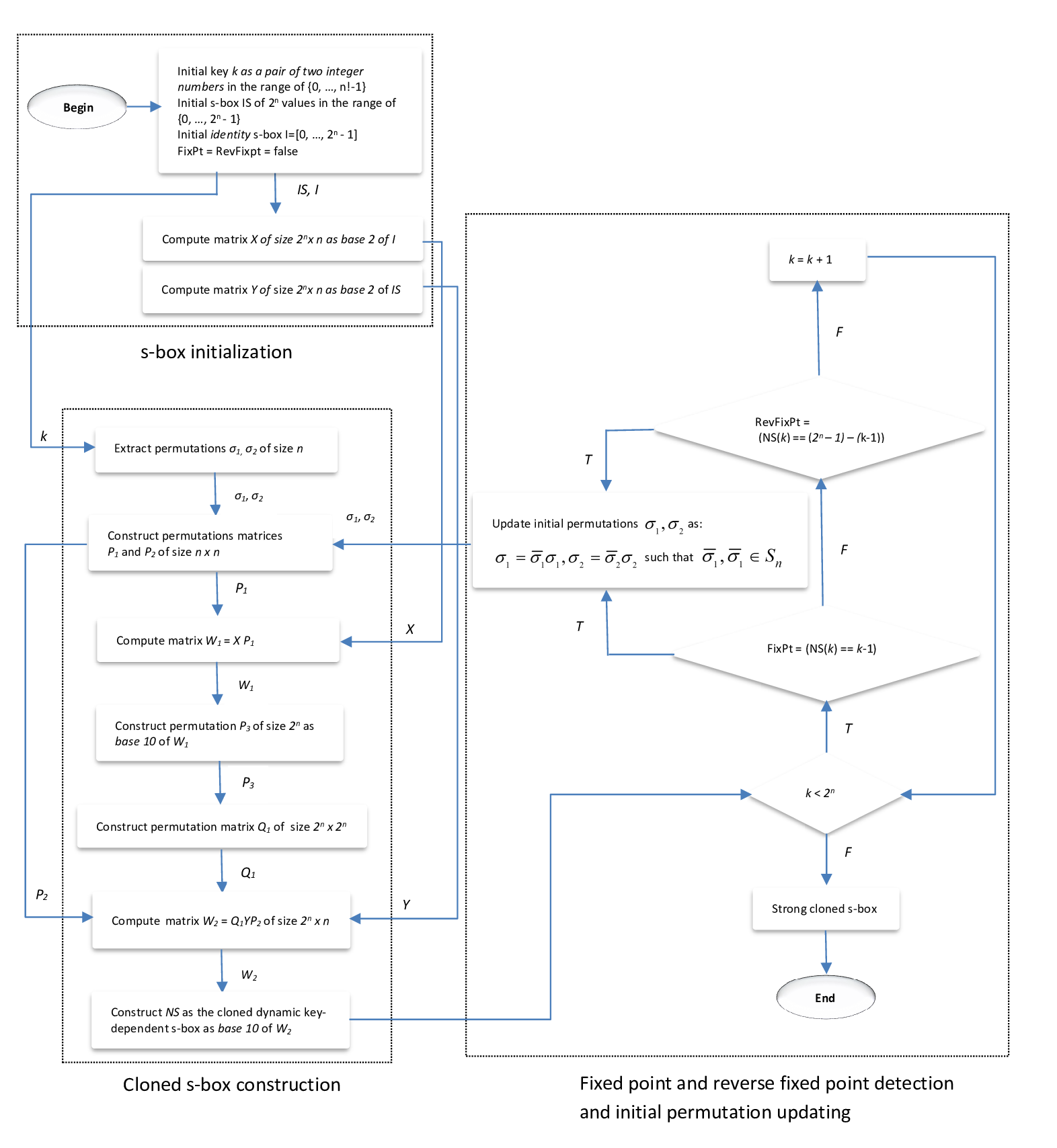}
\caption{Flowchart of constructing cloned key-dependent s-box}
\end{figure}
\section{Performance analysis}
This section provides demonstration of how our algorithm can be applied to construct clone copies for a given s-box while preserving its cryptographic features and strength. The application is independent of the method used for construction of the given s-box. In case of the initial given s-box having fixed points or reverse fixed points, the algorithm can be applied to obtain improved clone versions where all the fixed points and reverse fixed points are removed but the specifications like bijection, nonlinearity, SAC, and BIC are conserved with same strength as the initial given s-box. This adds particular significance and increases the scope of applications of the algorithm in the context of the recent analysis \cite{Liu2020} of the exploitable weakness of fixed point and reverse fixed point contained in s-boxes.

The performance of our method is illustrated through the following two examples:
\begin{example}{Demonstration of algorithm for $n=4$}

We use  the initial $4\times 4$ s-box given as a vector $IS$
\begin{equation}\label{IS}
IS =[9, 13, 10, 15, 11, 14, 7, 3, 12, 8, 6, 2, 4, 1, 0, 5]^t
\end{equation}
constructed by Carlisle Adams and Stafford Tavares
\cite{Adms1990}. Let us assume that the key gives the  two
permutations $\sigma_1=(1,2,0,3)$, $\sigma_2=(3,2,0,1)$ of size
$4$.

The corresponding $X,Y,P_1,P_2, W_1,P_3, Q_1, W_2$ are found as
$$X=\left[
\begin {array}{cccccccccccccccc}
0&1&0&1&0&1&0&1&0&1&0&1&0&1&0&1\\
0&0&1&1&0&0&1&1&0&0&1&1&0&0&1&1\\
0&0&0&0&1&1&1&1&0&0&0&0&1&1&1&1\\
0&0&0&0&0&0&0&0&1&1&1&1&1&1&1&1
\end {array}
\right]^t
$$

$$ Y=\left[
\begin {array}{cccccccccccccccc}
1&1&0&1&1&0&1&1&0&0&0&0&0&1&0&1\\
0&0&1&1&1&1&1&1&0&0&1&1&0&0&0&0\\
0&1&0&1&0&1&1&0&1&0&1&0&1&0&0&1 \\
1&1&1&1&1&1&0&0&1&1&0&0&0&0&0&0
\end {array}
\right]^t
$$

$P_1=\left[
\begin {array}{cccc}
 0&1&0&0\\
0&0&1&0\\
1&0&0&0\\
0&0&0&1
\end {array}
\right],
P_2=\left[
\begin {array}{cccc}
0&0&0&1\\
0&0&1&0\\
1&0&0&0\\
0&1&0&0
\end {array}
\right] $

$W_1= \left[
\begin {array}{cccccccccccccccc}
0&0&0&0&1&1&1&1&0&0&0&0&1&1&1 &1\\
0&1&0&1&0&1&0&1&0&1&0&1&0&1&0&1\\
0&0&1&1&0&0&1&1&0&0&1&1&0&0&1&1\\
0&0&0&0&0&0&0&0&1&1&1&1&1&1&1&1
\end {array}
\right]^t$

$P_3=(0,2,4,6,1,3,5,7,8,10,12,14,9,11,13,15)$

$Q_1=\left[ \begin {array}{cccccccccccccccc}
1&0&0&0&0&0&0&0&0&0&0&0&0&0&0 &0\\
0&0&1&0&0&0&0&0&0&0&0&0&0&0&0&0\\
0&0&0&0&1&0&0&0&0&0&0&0&0&0&0&0\\
0&0&0&0&0&0&1&0&0&0&0&0&0&0&0&0\\
0&1&0&0&0&0&0&0&0&0&0&0&0&0&0&0\\
0&0&0&1&0&0&0&0&0&0&0&0&0&0&0&0\\
0&0&0&0&0&1&0&0&0&0&0&0&0&0&0&0\\
0&0&0&0&0&0&0&1&0&0&0&0&0&0&0&0\\
0&0&0&0&0&0&0&0&1&0&0&0&0&0&0&0\\
0&0&0&0&0&0&0&0&0&0&1&0&0&0&0&0\\
0&0&0&0&0&0&0&0&0&0&0&0&1&0&0&0\\
0&0&0&0&0&0&0&0&0&0&0&0&0&0&1&0\\
0&0&0&0&0&0&0&0&0&1&0&0&0&0&0&0\\
0&0&0&0&0&0&0&0&0&0&0&1&0&0&0&0\\
0&0&0&0&0&0&0&0&0&0&0&0&0&1&0&0\\
0&0&0&0&0&0&0&0&0&0&0&0&0&0&0&1\\
\end {array}
\right] $

$W_2=\left[
\begin {array}{cccccccccccccccc}
0&0&0&1&1&1&1&0&1&1&1&0&0&0&0 &1\\
1&1&1&0&1&1&1&0&1&0&0&0&1&0&0&0\\
0&1&1&1&0&1&1&1&0&1&0&0&0&1&0&0\\
1&0&1&1&1&1&0&1&0&0&0&0&0&0&1&1
\end {array}
\right]^t$

Finally, the vector $NS$
 \begin{equation}\label{NS}
NS=[10,6,14,13,11,15,7,12,3,5,1,0,2,4,8,9]^t
\end{equation}
provides the dynamic key-dependent s-box, having the same four
algebraic properties as the initial vector $IS$ as shown in the following table.
\begin{table}[h]
	\centering
	\resizebox*{1.00\textwidth}{!}{
		\begin{tabular}{|c|ccc|cccc|cccc|cccc|}
			\hline
			\multirow{2}{*}{}  & \multicolumn{3}{c|}{\textbf{Nonlinearity}} & \multicolumn{4}{c|}{\textbf{SAC}} &  \multicolumn{4}{c|}{\textbf{BIC of nonlinearity }}&  \multicolumn{4}{c|}{\textbf{BIC of SAC }}  \\
		    & \textbf{min} & \textbf{max} & \textbf{avg} & \textbf{min} & \textbf{max} & \textbf{avg}& \textbf{SD} & \textbf{min} & \textbf{max}&\textbf{avg} & \textbf{SD} & \textbf{min}& \textbf{max} &\textbf{avg} & \textbf{SD}   \\ \hline
				Initial  s-box  IS  &  4 &  4 &  4 & 0  & 1  &  0.5 & 0.132583   & 4  & 4  &  4 & 0  &  0.4375 & 0.75  &  0.552083  & 0.104686     \\
				Clone copy s-box NS  &  4 &  4 &  4 & 0  & 1  &  0.5 & 0.132583   & 4  & 4  &  4 & 0  &  0.4375 & 0.75  &  0.552083  & 0.104686     \\ \hline
		\end{tabular}
	}
	\caption{Comparison of the algebraic properties of the initial  s-box IS (Equation \ref{IS}) and its clone copy NS (Equation  \ref{NS}) resulting from applying  the two
permutations $\sigma_1=(1,2,0,3)$, $\sigma_2=(3,2,0,1)$  on IS}
	\label{Comparison0}
\end{table}
\end{example}
\newpage
\begin{remark}
The algorithm was applied using the initial $4\times4$ s-box and
all the $4!$ permutations of size $4$. As a result, $(4!)^2=576$
different s-boxes were generated and it was verified that all have
the same four algebraic properties as the initial s-box.
\end{remark}
\begin{example}{Demonstration of algorithm for $n=8$}

We use  the initial $8\times 8$ AES s-box  constructed by  Joan Daemen and Vincent Rijmen
 \cite{daemen1991design} given in  Table \ref{AES}. Let us assume that the key gives the  two
permutations $\sigma_1=(1, 2, 0, 6, 5, 7, 3, 4)$, $\sigma_2=(5, 7, 3, 4, 1, 2, 0, 6)$ of size
$8$. Simillarly, applying the new method provides new s-box given in Table  \ref{clone AES} with the same four
algebraic properties as the initial AES s-box.
\begin{table}[h]
	\centering
	\resizebox*{.8\textwidth}{!}{
		\begin{tabular}{c c c c c c c c c c c c c c c c c} 
			\hline
                   R/C& 0 & 1 & 2 & 3 &4 & 5 & 6& 7 & 8 & 9 & 10 & 11& 12 & 13 & 14 & 15 \\
			\hline
                    \hline
                 &99& 124& 119& 123& 242& 107& 111& 197& 48& 1& 103& 43& 254& 215& 171& 118\\
                 &202& 130& 201& 125& 250& 89& 71& 240& 173& 212& 162& 175& 156& 164& 114& 192\\
                  &183& 253& 147& 38& 54& 63& 247& 204& 52& 165& 229& 241& 113& 216& 49& 21\\
                  &4& 199& 35& 195& 24& 150& 5& 154& 7& 18& 128& 226& 235& 39& 178& 117\\
                  &9& 131& 44& 26& 27& 110& 90& 160& 82& 59& 214& 179& 41& 227& 47& 132\\
                  &83& 209& 0& 237& 32& 252& 177& 91& 106& 203& 190& 57& 74& 76& 88& 207\\
                  &208& 239& 170& 251& 67& 77& 51& 133& 69& 249& 2& 127& 80& 60& 159& 168\\
                  &81& 163& 64& 143& 146& 157& 56& 245& 188& 182& 218& 33& 16& 255& 243& 210\\
                  &205& 12& 19& 236& 95& 151& 68& 23& 196& 167& 126& 61& 100& 93& 25& 115\\
                  &96& 129& 79& 220& 34& 42& 144& 136& 70& 238& 184& 20& 222& 94& 11& 219\\
                  &224& 50& 58& 10& 73& 6& 36& 92& 194& 211& 172& 98& 145& 149& 228& 121\\
                  &231& 200& 55& 109& 141& 213& 78& 169& 108& 86& 244& 234& 101& 122& 174& 8\\
                  &186& 120& 37& 46& 28& 166& 180& 198& 232& 221& 116& 31& 75& 189& 139& 138\\
                  &112& 62& 181& 102& 72& 3& 246& 14& 97& 53& 87& 185& 134& 193& 29& 158\\
                  &225& 248& 152& 17& 105& 217& 142& 148& 155& 30& 135& 233& 206& 85& 40& 223\\
                  &140& 161& 137& 13& 191& 230& 66& 104& 65& 153& 45& 15& 176& 84& 187& 22\\
			\hline
			\hline
	\end{tabular}}
	\caption{Presentation of  AES s-box \cite{daemen1991design} in $16 \times16$ matrix form}
	\label{AES}
\end{table}
\begin{table}[h]
\centering
\resizebox*{.8\textwidth}{!}{
\begin{tabular}{c c c c c c c c c c c c c c c c c} 
\hline
 0 & 1 & 2 & 3 &4 & 5 & 6& 7 & 8 & 9 & 10 & 11& 12 & 13 & 14 & 15 \\
\hline
\hline
165&175&199&189&31&183&181&
105&48&28&178&147&224&146&157&68\\ \noalign{\medskip}238&226&142&239&
127&140&190&89&67&212&161&166&253&247&57&104\\ \noalign{\medskip}121&
162&187&9&24&93&234&170&214&44&26&78&23&156&204&201
\\ \noalign{\medskip}69&150&49&12&134&144&136&27&101&82&53&216&87&34&
115&74\\ \noalign{\medskip}6&173&223&244&32&180&235&143&131&203&52&188
&182&230&229&72\\ \noalign{\medskip}14&109&39&38&108&103&83&42&41&128&
3&250&119&191&30&84\\ \noalign{\medskip}73&159&13&50&236&62&59&167&85&
15&177&240&123&186&126&208\\ \noalign{\medskip}193&92&98&77&227&133&
106&55&242&232&217&20&154&117&43&251\\ \noalign{\medskip}209&113&215&
169&192&63&51&71&163&0&4&102&99&125&95&179\\ \noalign{\medskip}8&164&
18&40&233&225&202&210&35&1&194&22&228&248&122&111\\ \noalign{\medskip}
5&185&132&66&96&91&148&80&7&110&17&207&158&141&160&152
\\ \noalign{\medskip}237&174&120&153&81&61&107&116&88&112&254&129&100&
56&205&21\\ \noalign{\medskip}124&196&90&135&75&252&76&65&149&222&145&
19&241&54&25&249\\ \noalign{\medskip}168&64&245&198&130&197&172&47&94&
211&2&231&206&36&255&195\\ \noalign{\medskip}137&86&219&176&221&10&155
&243&37&171&200&58&46&118&97&218\\ \noalign{\medskip}29&79&45&220&139&
213&151&16&33&60&70&246&114&184&11&138\\
\hline
\hline
\end{tabular}}
\caption{Presentation of clone copy s-box  as $16 \times16$ matrix resulting from applying  the two
permutations $\sigma_1=(1, 2, 0, 6, 5, 7, 3, 4)$, $\sigma_2=(5, 7, 3, 4, 1, 2, 0, 6)$  on AES s-box shown in Table \ref{AES}}
\label{clone AES}
\end{table}
\begin{table}[h]
	\centering
	\resizebox*{1.00\textwidth}{!}{
		\begin{tabular}{|c|ccc|cccc|cccc|cccc|}
			\hline
			\multirow{2}{*}{}  & \multicolumn{3}{c|}{\textbf{Nonlinearity}} & \multicolumn{4}{c|}{\textbf{SAC}} &  \multicolumn{4}{c|}{\textbf{BIC of nonlinearity }}&  \multicolumn{4}{c|}{\textbf{BIC of SAC }}  \\
		    & \textbf{min} & \textbf{max} & \textbf{avg} & \textbf{min} & \textbf{max} & \textbf{avg}& \textbf{SD} & \textbf{min} & \textbf{max}&\textbf{avg} & \textbf{SD} & \textbf{min}& \textbf{max} &\textbf{avg} & \textbf{SD}   \\ \hline
				AES s-box    &      112 & 112 & 112 &  0.453125 & 0.5625  &  0.504883  & 0.015678& 112 &112 & 112& 0 &0.480469&0.525391&0.504604&0.011271 \\
				Clone copy of AES s-box  &  112 & 112 & 112 &  0.453125 & 0.5625  &  0.504883  & 0.015678& 112 &112 & 112& 0 &0.480469&0.525391&0.504604&0.011271 \\ \hline
		\end{tabular}
	}
	\caption{Comparison of the algebraic properties of AES s-box (Table \ref{AES}) and its clone copy (Table  \ref{clone AES})}
	\label{Comparison}
\end{table}

\end{example}
\newpage
\section{Conclusion}
This work investigates the question of generating key-dependent dynamic $n\times n$ clone s-boxes having the same algebraic properties. Using initial s-box, we provide an algorithmic approach to generate clone s-boxes which have the same genetic traits like bijection, nonlinearity, SAC, and BIC. Invariance of the bijection, nonlinearity, SAC, and BIC for the generated clone copies is proved. The flow chart and Maple code of the presented algorithm are also given. The efficiency of the algorithm is tested through examples.  In conclusion, instead of focusing on finding ways to generate strong s-boxes, it may be enough to start with one strong s-box such as AES s-box, S8 AES s-box,  APA s-box, and Gray s-box and then get its clone copies.
\subsection*{Acknowledgments}
 The authors would like to thank Qatar University for its support and excellent research
facilities.
\section*{Appendix: Maple code for the proposed algorithm}
\lstset{basicstyle=\scriptsize}
\begin{lstlisting}[caption={Maple Procedure for converting a permutation $\sigma$ to a permutation matrix $R$},frame=single]
permatrix:=proc(sigma)
local R,i;
R:=Matrix(nops(sigma),nops(sigma),[0]);
for i from 1 to nops(sigma) do R[i,sigma[i]+1]:=1; end do;
R;
end proc:
\end{lstlisting}
\lstset{basicstyle=\scriptsize}
\begin{lstlisting}[caption={Maple Procedure for  evaluating the binary representation of  an s-box $S$ as a Boolean matrix  $M$ of size $2^n \times n$  },frame=single]
BLmatrix := proc(S,n)
local M;
M:=Matrix([seq(convert([S[i]],base,10,2),i=1..nops(S))]);
end proc:
\end{lstlisting}
\lstset{basicstyle=\scriptsize}
\begin{lstlisting}[caption={Maple Procedure for  evaluating the decimal representation of  a Boolean matrix  $M$ of size $2^n \times n$ as an s-box $S$},frame=single]
Sbox := proc(M,n)                                                                                                                                                                     
local S;                                                                                                                   
S:=[seq(add(convert(Row(M,i),list)[j]*2^(j-1),j=1..n),i=1..RowDimension(M))];                                               
end proc:
\end{lstlisting}
\lstset{basicstyle=\scriptsize}
\begin{lstlisting}[caption={Maple code for the proposed algorithm for constructing new s-box $NS$ using the initial s-box $IS$ and the two permutations $\sigma_1$ and $\sigma_2$ which extracted from the key},frame=single]
NS := proc(IS,sigma1,sigma2)  
local n,Y,IDSBox,X,P1,P2,W1,P3,Q1,W2,NS;
n:=log[2](nops(IS));
Y:=BLmatrix(IS,n);
IDSBox:=[seq(i,i=0..2^n-1)]; X:=BLmatrix(IDSBox,n);
P1:=permatrix(sigma1); P2:=permatrix(sigma2);
W1:=X.P1; P3:=Sbox(W1,n); Q1:=permatrix(P3);
W2:=Q1.Y.P2; 
NS:=Sbox(W2,n)
end proc:
\end{lstlisting}
\lstset{basicstyle=\scriptsize}
\begin{lstlisting}[caption={Maple code for deduction of fixed point and reverse fixed point for s-box $S$ },frame=single]
FIXP :=proc(S)                                                                                                                                                                 local n,i,FPS,RFPS;                                                                                                                                                  n:=nops(S);                                                                                                                                                                    FPS:={};                                                                                                                                                                     RFPS:={};                                                                                                                                                                          for i from 1  to n do                                                                                                                                                           if S[i]=i-1 then                                                                                                                                                  
FPS:={S[i]}union  FPS;                                                                                                                                                      elif  S[i]=255-(i-1) then                                                                                                                                               RFPS:={S[i]}union  RFPS;                                                                                                                                                  end if;                                                                                                                                                                             end do;                                                                                                                                                             [FPS,RFPS];                                                                                                                                                                   end proc:
\end{lstlisting}
Conflict of Interest: The authors declare that they have no conflict of interest.


\begin{thebibliography}{99}
\bibitem{Liu2020}Liu, Hongjun, Abdurahman Kadir, and Chengbo Xu. "Cryptanalysis and constructing S-Box based on chaotic map and backtracking." Applied Mathematics and Computation 376 (2020): 125153.
\bibitem{27} Easttom C. An Examination of Inefficiencies in Key Dependent Variations of the Rijndael S-Box. InElectrical Engineering (ICEE), Iranian Conference on 2018 May 8 (pp. 1658-1663). IEEE.
\bibitem{n15}Iqtadar Hussain, Tariq Shah Hasan Mahmood, Muhammad Asif Gondal. Construction of S8 Liu J S-boxes and their applications. Computers and Mathematics with Applications 64 (2012) 2450–2458.
\bibitem{n16}Iqtadar Hussain, Tariq Shah, Muhammad Asif Gondal, Hasan Mahmood. S8 affine-power-affine S-boxes and their applications. Neural Comput and Applic (2012) 21 (Suppl 1):S377–S383.
\bibitem{n30} Iqtadar Hussain, Amir Anees, Muhammad Aslam, Rehan Ahmed, Nasir Siddiqui. A noise resistant symmetric key cryptosystem based on S8 S-boxes and chaotic maps. The European Physical Journal Plus, April 2018, pp.133:167. 
\bibitem{n32} Designing secure substitution boxes based on permutation of symmetric group Amir Anees, Yi-Ping Phoebe Chen. Neural Computing and Applications. https://doi.org/10.1007/s00521-019-04207-8
\bibitem{Adms1990} Adms, C.M., Tavares, S.E., 1990. The Structured Design of Cryptographically Good S-Boxes, Journal of Cryptology, 3 (1), pp. 27-41.
\bibitem{1} Musheer Ahmad, Parvez  Khan, Mohd Ansari, A Simple and Efficient Key-Dependent SBox Design Using Fisher-Yates Shuffle Technique, International Conference on Security in Computer Networks and Distributed Systems SNDS 2014: Springer, Recent Trends in Computer Networks and Distributed Systems Security pp 540-550.
\bibitem{2} M. Dawson, S. Tavares, An expanded set of design criteria for substitution boxes and their use in strengthening DES-like cryptosystems, In IEEE Pacific rim conference on communications, computers and signal processing, 913 May 1991. pp. 1915.
\bibitem{3} YongWang, Kwok-WoWong, Xiaofeng Liao, Tao Xiang, A block cipher with dynamic Sboxes based on tent map, Communications in Nonlinear Science and Numerical Simulation, 2009(14):3089-3099.
\bibitem{4} G.N. Krishnamurthy and V. Ramaswamy, Making AES Stronger: AES with Key Dependent S-Box, International Journal of Computer Science and Network Security, vol. 8, pp. 388398, 2008.
\bibitem{5} M. Piotr, Generating Pseudorandom S-Boxes a Method of Improving the Security of Cryptosystems Based on Block Ciphers, Journal of Telecommunications and Information Technology, 2009.
\bibitem{6} N. Stoianov, "One approach of using key-dependent S-BOXes in AES," in Multimedia Communications, Services and Security, ed: Springer, 2011, pp. 317-323.
\bibitem{7} A. ElGhafar, A. Rohiem, A. Diaa and F. Mohammed, Generation of AES Key Dependent SBoxes using RC4 Algorithm, 13th International Conference on Aerospace Sciences Aviation Technology, ASAT- 13, 2009, pp. 26-28.
\bibitem{8} K. Kazys and K. Jaunius, Key-Dependent S-Box Generation in AES Block Cipher System, INFORMATICA, 2009, pp. 23-34.
\bibitem{9} Z. Ghada, K. Abdennaceur, P. Fabrice and F. Daniele, On Dynamic chaotic S-BOX, IEEE, 2009.
\bibitem{10} J. Cui, L. Huang, H. Zhong, C. Chang and W. Yang, An Improved AES S-Box and Its Performance Analysis, International Journal of Innovative Computing, Information and Control, 2011.
\bibitem{11} G. Anna, Cryptographic properties of modified AES-like S-boxes, Annales UMCS Informatica AI XI, 2011; 2, 37-48.
\bibitem{12} J. Julia, M. Ramlan, S. Salasiah and R. Jazrin, Enhancing Advanced Encryption Standard SBox Generation Based on Round Key,, IJCSDF, vol. 3, pp. 183-188, 2012.
\bibitem{13} R. Hosseinkhani and H. Haj Seyyed Javadi, Using Cipher Key to Generate Dynamic S-Box in AES Cipher System, IJCSS, vol. 6, pp. 19-28, 2012.
\bibitem{14} O. Kazymyrov, V. Kazymyrova and R. Oliynykov, A Method For Generation Of HighNonlinear S-Boxes Based On Gradient Descent, IACR Cryptology, 2013.
\bibitem{15} M. Dara and K. Manochehri, A Novel Method for Designing S-Boxes Based on Chaotic Logistic Maps Using Cipher Key, World Applied Sciences Journal, vol. 28, pp. 2003-2009, 2013.
\bibitem{16} E. Mohammed Mahmoud , A. Abd El Hafez, A. Talaat and A. Zekry, Dynamic AES-128 with key-dependent s-box, International Journal of Engineering Research and Applications, vol. 3, 1662-1670, 2013.
\bibitem{17} S. Arrag, A. Hamdoun, A. Tragha and S. Eddine Khamlich, Implementation Of Stronger AES By Using Dynamic S-Box Dependent Of Master Key, Journal of Theoretical and Applied Information Technology. 2013.
\bibitem{18} F. Ahmed and D. Elkamchouchi, Strongest AES with S-Boxes Bank and Dynamic Key MDS Matrix (SDK-AES), International Journal of Computer and Communication Engineering, 2013.
\bibitem{19} K. Adi Narayana Reddy and B. Vishnuvardhan, Secure Linear Transformation Based Cryptosystem using Dynamic Byte Substitution, International Journal of Security, 2014.
\bibitem{20} K. Kazys, V. Gytis and S. Robertas, An Algorithm for Key-Dependent S-Box Generation in Block Cipher System, INFORMATICA, vol. 26, pp. 51-65, 2015.
\bibitem{21} K. Balajee Maram and J.M. Gnanasekar, Evaluation of Key Dependent S-Box Based Data Security Algorithm using Hamming Distance and Balanced Output, TEM Journal. 2016.
\bibitem{22} S. Katiyar and N. Jeyanthi, Pure Dynamic S-box Construction, International Journal of Computers, 2016.
\bibitem{23} A. Tianyong, R. Jinli, D. Kui, and Z. Xuecheng, Construction of High Quality Keydependent S-Box, IAENG International Journal of Computer Science, 2017.
\bibitem{24} C. Unal, K. Sezgin, P. Ihsan and Z. Ahmet, Secure image encryption algorithm design using a novel chaos based S-Box, Chaos, Solitons and Fractals, vol. 95, pp. 92-101, 2017.
\bibitem{25} P. Agarwal, A. Singh and A. Kilicman, Development of Key Dependent Dynamic S-Boxes with Dynamic Irreducible Polynomial and Affine Constant, Advances in Mechanical Engineering, 2018.
\bibitem{26} A. Singh, P. Agarwal, M. Chand, Image Encryption and Analysis using Dynamic AES, 5th International Conference on Optimization and Applications (ICOA), 2019.
\bibitem{Shannon1949} Shannon, C. E., 1949. Communication theory of secrecy system, The Bell System Technical Journal, 28: pp. 656715.
\bibitem{Pieprzyk1988}J. Pieprzyk and G. Finkelstein, Towards effective nonlinear cryptosystem design, IEE Proceedings, Part E: Computers and Digital Techniques, 135 (1988), 325-335.
\bibitem{Pieprzyk2013}Pieprzyk, J., Hardjono, T. and Seberry, J., 2013. Fundamentals of computer security. Springer Science and Business Media.
\bibitem{Webster1986} Webster, A. F., Tavares, S. E., 1986. On the Design of S-Boxes, In Proceedings of Advances in CryptologyCRYPTO 85. Lecture Notes in Computer Science, Vol. 218, Springer-Verlag, pp. 523-534.
\bibitem{daemen1991design} J. Daemen, V. Rijmen, AES proposal: Rijndael,\\
 http://csrc.nist.gov/archive/aes/rijndael/Rijndael-ammended.pdf (1999).
\end{thebibliography}
\end{document}